\documentclass[12pt]{article}

\usepackage{graphicx}
\usepackage{subfigure}

\begin{document} 
\begin{titlepage}       
\begin{center}
{\Large {\bf Two Geometric Approaches To Study The Deconfinement 
Phase Transition in (3+1)-Dimensional $Z_2$ Gauge Theories}}
\end{center}
\vskip 1.5cm
\centerline{\bf Semra G\"{u}nd\"{u}\c{c}, Mehmet Dilaver  and  Yi\u{g}it G\"{u}nd\"{u}\c{c}}

\centerline{\bf Hacettepe University, Physics Department,}
\centerline{\bf 06532  Beytepe, Ankara, Turkey }

\vskip 0.5cm

\centerline{\normalsize {\bf Abstract} }

We have simulated $(3+1)-$dimensional finite temperature $Z_2$ gauge
theory by using Metropolis algorithm. We aimed to observe the
deconfinement phase transitions by using geometric methods. In order
to do so we have proposed two different methods which can be applied
to $3-$dimensional effective spin model consisting of Polyakov loop
variables. The first method is based on studies of cluster structures
of each configuration. For each temperature, configurations are
obtained from a set of bond probability ($P$) values.  At a certain
probability, percolating clusters start to emerge. Unless the
probability value coincides with the Coniglio-Klein probability value,
the fluctuations are less than the actual fluctuations at the critical
point. In this method the task is to identify the probability value
which yields the highest peak in the diverging quantities on finite
lattices. The second method uses the scaling function based on the
surface renormalization which is of geometric origin. Since this
function is a scaling function, the measurements done on
different-size lattices yield the same value at the critical point,
apart from the correction to scaling terms.  The linearization of the
scaling function around the critical point yields the critical point
and the critical exponents.

\vskip 0.4cm


\end{titlepage} 

%\pagebreak
 
\section{Introduction}

Lattice studies of quantum chromodynamics (QCD) show that there exist
two different phase transitions at two extreme limits. The first one
is the infinitely heavy quark mass limit. This limit is relatively easy to
study since $QCD$ with static quarks can be studied as pure gauge
theory. In this limit, gauge theories undergo a phase transition from
a low temperature phase in which quarks and gluons are confined, to a
high temperature phase in which colored quarks and gluons exist. This
transition is named as the deconfinement phase transition and can be
studied in terms of its order parameter, Polyakov loop $(L)$~\cite{Polyakov-Susskind}, 
which is similar to the spin variables in
ferromagnetic systems. The deconfinement phase transition of $SU(N)$
and $Z_N$ gauge theories are well-understood in terms of Svetitsky and
Yaffe (S-Y) conjecture~\cite{SvetitskyYaffe}. According to S-Y
conjecture, $(d+1)-$dimensional finite temperature $SU(N)$ and $Z_N$
lattice gauge theories are in the same universality class as
$d-$dimensional $Z_N$ spin system. The problem appears when one wants
to study gauge theories with finite-mass quarks. In the finite
temperature lattice gauge theories, the quark fields couple to Polyakov
loops similarly to the coupling of external magnetic field to spin
variables in the ferromagnetic systems~\cite{Kertesz}. Hence the Polyakov loop is no
longer the order parameter for the deconfinement phase
transition. There have been some recent attempts to device new
observables for studying the deconfinement phase transition in the
finite quark mass limit~\cite{Satz1,Satz2}.  The importance of the
definition of a new order parameter is that it may enable one to study
the relation between the deconfinement and chiral phase transitions~\cite{Satz2}
which is seen in the second extreme limit where the quarks are
massless. It is known that quarks have finite and relatively low
mass. One task is to understand whether the deconfinement phase
transition continues into finite-mass region or not.

$\;$

The relation between ferromagnetic phase transition and the
percolation theory has been established some years ago by Coniglio and
Klein~\cite{KasteleynFortuin,ConiglioKlein}. One of the proposed
solutions to the above mentioned problem is based on the studies
of the percolating clusters existing in the system around the phase
transition region. This approach is used to introduce a geometric method
of following the deconfinement phase transition in lattice gauge
theories~\cite{Fort1,Fort2,Fort3,Fort4,Blanchard,Fort5,Bialas}.
The main difficulty in applying Coniglio-Klein approach to lattice
gauge theories is to obtain a finite effective spin Hamiltonian from
lattice gauge theory action.  In finite temperature lattice gauge
theories Polyakov loop is the unique gauge invariant quantity which is
sensitive to symmetry breaking. From $(d+1)-$dimensional finite
temperature lattice gauge theory, one can obtain a $d-$dimensional
effective spin system whose dynamic variables are Polyakov 
loops~\cite{SvetitskyYaffe,GreenKarsch}.  In this work
we aim to apply two different techniques to the effective spin system
obtained from the finite temperature gauge theory.  The success of
these techniques are previously tested on spin systems for first- and
second-order phase 
transitions~\cite{semraprob,oliveira1,oliveira2,oliveira3,oliveira4,semrascaling}.

$\;$

The first approach which we will apply to the $(3+1)-$dimensional finite
temperature Ising gauge theory is the method which is based on
determining the Coniglio-Klein-like probability value which produces
the largest fluctuations in the $3-$dimensional effective spin system
consisting of the Polyakov loop variables~\cite{semraprob}.  The
second approach that we want to implement on lattice gauge theories
uses the scaling function based on the surface
renormalization~\cite{oliveira1,oliveira2,oliveira3,oliveira4,semrascaling}.
Both methods can easily be implemented to follow the deconfinement
phase transition into finite mass quark region.

$\;$

The paper is organized as follows. In Section \ref{MM} we explain how
to implement the approaches to finite temperature lattice gauge
theories. In Section \ref{RD} we have presented our simulation results
and finally, Section \ref{Conc} is devoted to the conclusions and
further possible applications of the methods.

\section{Model and the Method}
\label{MM}

The finite temperature on the lattice is defined by introducing finite
and periodic temporal direction, while in all $d$ spatial directions,
the lattice is infinite.  For computer simulations this can be
implemented by taking the spatial extent $(N_{\sigma})$ of the lattice
much larger than the temporal extent $(N_{\tau})$. In this work in
order to discuss the operators of geometric origin to study
deconfinement phase transition, we have considered the simplest possible
gauge theory, namely $Z_2$ lattice gauge theory. During our simulations
the temporal extent of the lattice is fixed as $N_\tau=2$.
The Hamiltonian of the Ising gauge theory is

\begin{equation}
H ={\sum_{i,j,\mu,\nu}}{U_{\mu}}(i){U_{\nu}}(i+\mu){U_{\mu}}^{\dagger}(j+{\mu}){U_{\nu}}^{\dagger}(j)
\end{equation}

where ${U_{\mu}}(i)$ represents gauge field which takes only two 
values, ${U_{\mu}}(i)={\pm 1}$, in this model.  The Polyakov loop variables 
are given by

\begin{equation}
L = {\Large {\prod}}_{i = 1}^{N_{\tau}}{U_{\tau}}(i).
\end{equation}

For each spatial lattice site, one can assign a Polyakov loop
variable. This $3$-dimensional reduced system can be considered as a
spin system.  For finite temperature $Z_2$ gauge theory this
$3-$dimensional spin system has dynamic variables $L={\pm}1$.

$\;$

In this work we propose to use two different methods with geometric
origin. The first method is based on the introduction of a set of
probability values which will determine the Coniglio-Klein-like
clusters for each inverse temperature $\beta$~\cite{semraprob}. The probability value,
which coincides with the actual Coniglio-Klein probability at the
transition temperature of the effective spin system, can be determined
by studying the highest fluctuations in the system. The largest
fluctuations are the indication of the phase transition~\cite{semraprob}.
Hence, one expects that the thermodynamic quantities
obtained by using this probability value obey the scaling rules
giving the critical exponents of the system.

$\;$

For completeness we will give a brief description of this method. In a
ferromagnetic system each configuration contains clusters of like
spins. These clusters percolate before the phase transition point and
it is known that these geometric clusters are not sufficient to
establish the connection between the percolation of these geometric
structures and thermal phase transitions. By using Fortuin-Kasteleyn
clusters~\cite{KasteleynFortuin}, Coniglio and
Klein~\cite{ConiglioKlein} showed that the partition function of the
Ising Model can be written in purely geometric terms. In this
description of the system one uses a bond probability value to define
the number of spins which belong to the same cluster. For the Ising
Model this bond probability for like spins is given as
$P=1-e^{-2J/kT}$.  This probability reduces the existing
geometric clusters of like spins in each configuration to smaller
clusters which are called Coniglio-Klein clusters. It is shown that in
all dimensions, the thermal phase transitions can be studied in terms
of percolating Coniglio-Klein clusters.  In order to obtain
Coniglio-Klein clusters, in other words, to find the correct bond
breaking probability, one needs to know the Hamiltonian of the
system. In the finite temperature lattice gauge theory case this
Hamiltonian is the Hamiltonian of the effective spin system.  Assuming
that the probability that breaks the existing geometric structures are
unknown. In this case, one can choose a set of trial probabilities
between 0 and 1 and decide which one does correspond to the correct
probability value.  Simulations are performed for a set of chosen
temperature values near the transition temperature. For each
temperature, a set of predetermined trial probability values may be
used to break the existing geometrical connections between like spins
at each configuration. In fact this range is not very wide since the
percolation threshold at each dimension is known for any given lattice
geometry~\cite{StaufferAharony}. Moreover a simple test run can limit
the range to obtain required accuracy.  If the original configuration
has a spanning cluster of like spins, a subset of probability values
may result in spanning clusters of Coniglio-Klein type. As one uses
the bond-breaking probability $P$, the tuning of this parameter may
point out the connection between the dynamics of the system and this
artificial parameter. As this parameter approaches to the value given
by Coniglio and Klein at a given inverse temperature $\beta$, these
clusters coincide with the dynamically generated Coniglio-Klein
($C$-$K$) clusters and $P$ value can be identified by the largest
fluctuations for each $\beta$ . Particularly, the value of the
parameter $P$ which corresponds to $\beta_{c}(L)$ results in the most
widely fluctuating cluster sizes.  The fluctuations in the system can
be followed by considering any fluctuating quantity. In this work we
have chosen the fluctuations of the largest cluster in each
configuration as the fluctuating quantity.

$\;$

For a certain probability $P$ the average of the largest cluster is defined as,

\begin{equation}
< C_{max} >_P\; =\;
{1\over{N_{c}}}{\sum_{i}^{{N_{c}}}}{1\over{V}}{(C_{i}^{max}(P))}
\end{equation}

where $N_c$ is the number of configurations, $C_{i}^{max}(P)$ is the
largest cluster obtained by using the probability $P$ in $i^{th}$
configuration and $V$ is the volume of the system.  The fluctuations
($MCF(P)$) of the largest clusters is given as

\begin{equation}
MCF(P) = < C_{max}^2 >_P - < C_{max} >_{P}^2 .
\end{equation}

These observables have the characteristic scaling behavior. 
${< C_{max}>}_{P}$  scales like magnetization and $MCF(P)$ scales 
like susceptibility;

\begin{equation}
< C_{max} >_P\; =\;N_{\sigma}^{Y_{H}-d} \; {\cal C} (t N_{\sigma}^{Y_T}) 
\end{equation}
and
\begin{equation}
MCF(P)\; =\; N_{\sigma}^{2Y_{H}-d} \; {\cal MCF}(t N_{\sigma}^{Y_T}) 
\end{equation}
where $N_{\sigma}$ is the linear size and $d$ is the dimension of the system, $t$
is the reduced temperature, $Y_{H}$ and $Y_{T}$ are the magnetic and
thermal critical exponents respectively.
In this method for each simulation temperature, the fluctuations of
the largest cluster will be studied for a predetermined set of
probability values. The $P$ value which results in the highest
fluctuations for a given $\beta$ value is taken as the value of the
Coniglio-Klein probability value. Particularly the highest peak for 
the simulation temperatures indicates the transition temperature.

$\;$

The second method uses the scaling function $F$ based on the surface
renormalization. The function $F$ is studied in detail for the Ising 
Model~\cite{oliveira1,oliveira2,oliveira3} and $q$-state Potts
Model~\cite{oliveira4,semrascaling}.  In order to calculate this
function, one considers the direction of the majority of spins of two
parallel surfaces which are $ L/2 $ distance away from each other. In
the calculations, a counter is increased by one only if two opposing
surfaces have majority of the spins in the same direction. For the
Ising Model, the formulation of this scaling function is simple, since
the direction of the majority of the spins can be calculated as the sign of
the sum of the spins in a surface~\cite{oliveira3}. Hence for the a spin system consisting of 
Ising spins,  $F$ can be written in the form

\begin{equation}
\label{ScalingFunction}
 F = < {\rm sign}[{S}_{i}] {\rm sign}[{S}_{i+L/2}]>
\end{equation}

where $S_i$ is the sum of the spins in the $i^{\rm th}$ surface.
Scaling function  $F$  obeys  the scaling relation 

\begin{equation}
F(t,N_{\sigma} ) \sim f(t N_{\sigma}^{Y_T}) .
\end{equation}

If $F$ is expanded in Taylor
series near the critical point $(t\rightarrow 0)$ in the form

\begin{equation}
\label{LinearF}
F \sim f(0,N_{\sigma}) +{\partial{f}\over\partial{t}} N_{\sigma}^{Y_T} t + ...
\end{equation}

the function $F$  can be written in the linearized form.
By using Eq.(\ref{LinearF}) one can obtain the thermal critical index as
well as the phase transition point of the effective spin system.

$\;$

The next section is devoted to our simulation results.

\section{Results and Discussions}
\label{RD}

We have simulated $(3+1)-$dimensional Ising gauge theory by using
Metropolis algorithm. In our simulations we have discarded
$2\times10^4$ iterations for thermalization.  The errors are
calculated by using Jack-Knife analysis on the measurements of $10$ to
$20$ different runs of $10^5 \; {\rm to}\; 10^6$ iterations after
thermalization.  Calculations are repeated for $40$ $\beta$ values and
for $15$ $P$ values.  In our simulations the time extent of the
lattice is kept fixed, $N_{\tau}=2$.  We have used four different size
spatial lattices, $N_{\sigma}=10,12,16\;{\rm and}\;20$, but we have
presented the results for the largest three lattice sizes.

$\;$

We have presented our results in two subsections. In the first
subsection (Method I) the results of the computations of fluctuations
in the system with respect to varying probability values are
presented. In the second subsection (Method II) the results obtained
from scaling function based on the surface renormalization are given.

\subsection{Method I}

This method is based on studies of cluster structure of each
configuration by repeatedly restructuring the configurations obtained
from a set of bond probability $(P)$ values. For each lattice size, $15 \; P$
values are used in order to obtain Coniglio-Klein-like clusters.
In this way the criticality of the system is searched by studying
cluster distribution at each inverse temperature ${\beta}$.  For each
${\beta}$ value at a certain probability, percolating clusters start
to emerge. Nevertheless, unless the probability value coincides with the
$C-K$ probability value, the fluctuations are less than the actual
fluctuations at the critical point. The task is to identify the
probability value which yields the highest peak in the diverging
quantities on finite lattices.

$\;$

In Figure 1a we have presented the fluctuations ($MCF$) of maximum size
clusters on lattice $N_{\sigma}=20$. In this figure we have plotted
the maximum fluctuations obtained at different temperatures. Here one
can see that the fluctuations increase as the temperature approaches
to a certain value (expected finite-size critical value for the
effective spin system) and reaches to its peak. The temperature
corresponding to the peak value is considered as the transition
temperature. In Figure 1b three highest fluctuation peaks are plotted
for closer inspection.

$\;$

Figure 2a shows the highest fluctuation peaks for spatial sizes
$N_{\sigma}=12,16 \;{\rm and}\;20$.  In figure 2b we have presented
the scaling of cluster fluctuations. The position of the peak yields
the phase transition point $\beta_c(N_{\sigma})$ for finite lattice
size $N_{\sigma}$ and the peaks of the fluctuations scales like
susceptibility with the lattice size in the form
$N_{\sigma}^{2Y_H-d}$.  The ${\beta_c(N_{\sigma})}$ is obtained by
minimizing the distance between scaled cluster fluctuation curves. The
thermal and magnetic critical exponents obtained after this
minimization as $Y_H=2.498 \pm 0.002, Y_T=1.587 \pm 0.002$ and
$\beta_c(N_{\sigma}=12)=0.4256 \pm 0.0003$,\,
$\beta_c(N_{\sigma}=16)=0.4254 \pm 0.0002$,\,
$\beta_c(N_{\sigma}=20)=0.4253\pm 0.0002$.  The difference between the
inverse transition temperature $\beta_{t}(\infty)$ for the infinite
lattice and $\beta_{t}(N_{\sigma})$ obtained for different size
lattices is given by

\begin{equation}
\label{BetaInf}
\beta_{t}(N_{\sigma}) - \beta_{t}(\infty) \sim N_{\sigma}^{-Y_{T}}.
\end{equation}

By using Eq.(\ref{BetaInf}) and calculated values of
$\beta_c(N_{\sigma})$ and $Y_T$ we have calculated the infinite
lattice critical point as $\beta_c(\infty)=0.4251\pm 0.0001$.

\subsection{Method II}

The scaling function $F$ given by Eq.(\ref{ScalingFunction}) is also
based on geometric structures on the lattice. In Figure 3a we have
plotted the scaling function obtained from $3-$dimensional effective
spin model for three different-size lattices in the full $\beta$
range.  In Figure 3b we have presented the linearized scaling function
in Eq.(\ref{LinearF}) around the critical point for all three
different-size lattices. Since $F$ is a scaling function, different
size lattices yield the same value at the transition point apart
from the correction to scaling terms.  From the pairwise crossing and
the slopes we have obtained phase transition point $\beta_c$ and the
thermal critical exponent $Y_T$ using Eq.(\ref{LinearF}). 
Calculated critical exponents and the transition point values are
given in Table 1. These two methods yield similar results. Particularly, the
location of the transition point by using two different
methods are in good agreement.

\section{Conclusions} 
\label{Conc}

In this work we have aimed to demonstrate two methods to follow the
deconfinement phase transition into finite mass region of lattice
quantum chromodynamics.  Starting from the gauge theory action one can
obtain an effective spin Hamiltonian (at high temperature
limit)~\cite{GreenKarsch}.  By using this effective Hamiltonian, in
principle, one can calculate various quantities to search the phase
transition. Nevertheless using full effective action is both
cumbersome and subject to approximations. In the literature this
approach is pursued only for the pure gauge theory case and also spin
systems with external magnetic field.  Avoiding the complication of an
effective Hamiltonian one can still obtain accurate information on the
behavior of the system through the deconfinement phase
transition. None of the two methods described in the present work
requires the knowledge of the interactions between the Polyakov
loops. Moreover, since only spin-like dynamic variables are under
consideration, the interactions between matter fields and the gauge
fields do not effect the conclusions drawn by using these two methods.
For $(3+1)$-dimensional finite temperature Ising gauge theory, we have
obtained critical exponents $Y_H$ and $Y_T$ from Coniglio-Klein like
clusters and $Y_T$ by using scaling function $F$. The results obtained
by using both methods are accurate and satisfactory for conclusive
evidence for the universality class of the phase transition.

\pagebreak

\section*{Acknowledgements}
We  thank M. Ayd{\i}n for discussions. 
We also greatfully acknowledge Hacettepe University Research Fund 
(Project no : 01 01 602 019) and Hewlett-Packard's Philanthropy Programme.

\pagebreak

\pagebreak

\section*{Table Captions} 
\begin{description}

\item  Table 1. The critical exponents and the transition point 
values obtained by using $F$ function.
\end{description}

\pagebreak

\section*{Figure Captions} 
\begin{description}

\item Figure 1.  The fluctuations $MCF$ of maximum size
clusters on lattice $N_{\sigma}=20$. a) for all $P_{max}(\beta)$, b) for three 
$P_{max}(\beta)$ values with highest peaks.

\item Figure 2. a) Highest fluctuation peaks for spatial sizes
$N_{\sigma}=12,16 \;{\rm and}\;20$.  b) The scaling of cluster fluctuations given in a).

\item Figure 3. a) The scaling function $F$ for three different size lattices 
$N_{\sigma}=12,16 \;{\rm and}\;20$. b) Linearized scaling function $F$ 
around the critical point for the same lattices.

\end{description}

\pagebreak

\begin{table}
\begin{center}
\begin{tabular}{|p{2cm}|p{3cm}|p{3cm}|}
\hline
$N_{\sigma_1}-N_{\sigma_2}$ &$Y_T$ &$\beta$\\
\hline
$12-16$ &$1.568 \pm 0.002$  & $0.4251 \pm 0.0002$\\
\hline
$16-20$&$1.666 \pm 0.006$  &$0.4251 \pm 0.0002 $ \\
\hline
\end{tabular}
\end{center}
\caption{}
\label{Table1}
\end{table}

\pagebreak

\begin{figure}
\centering
\subfigure[]{\includegraphics[angle=0,width=10truecm]{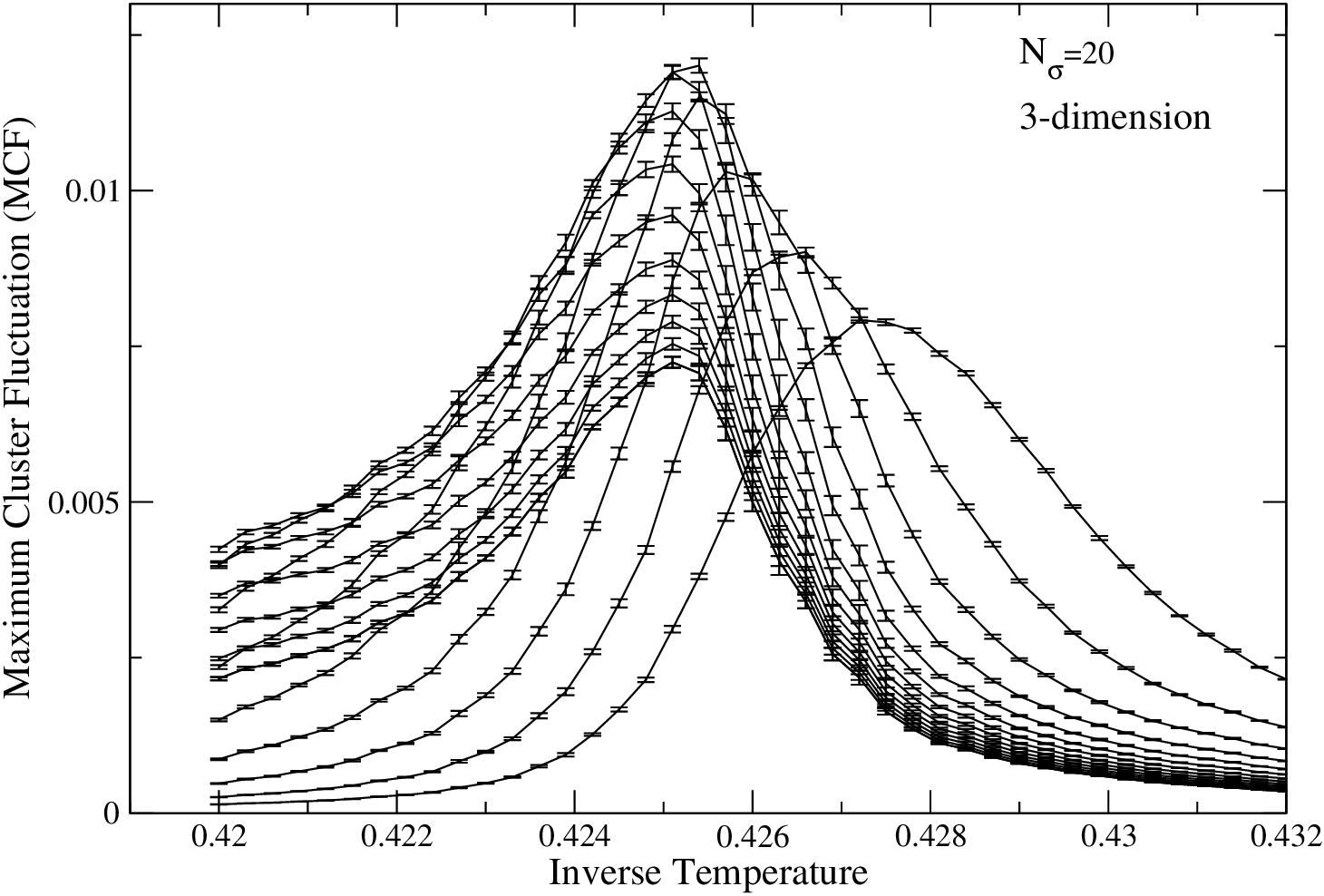}}\\
\subfigure[]{\includegraphics[angle=0,width=10truecm]{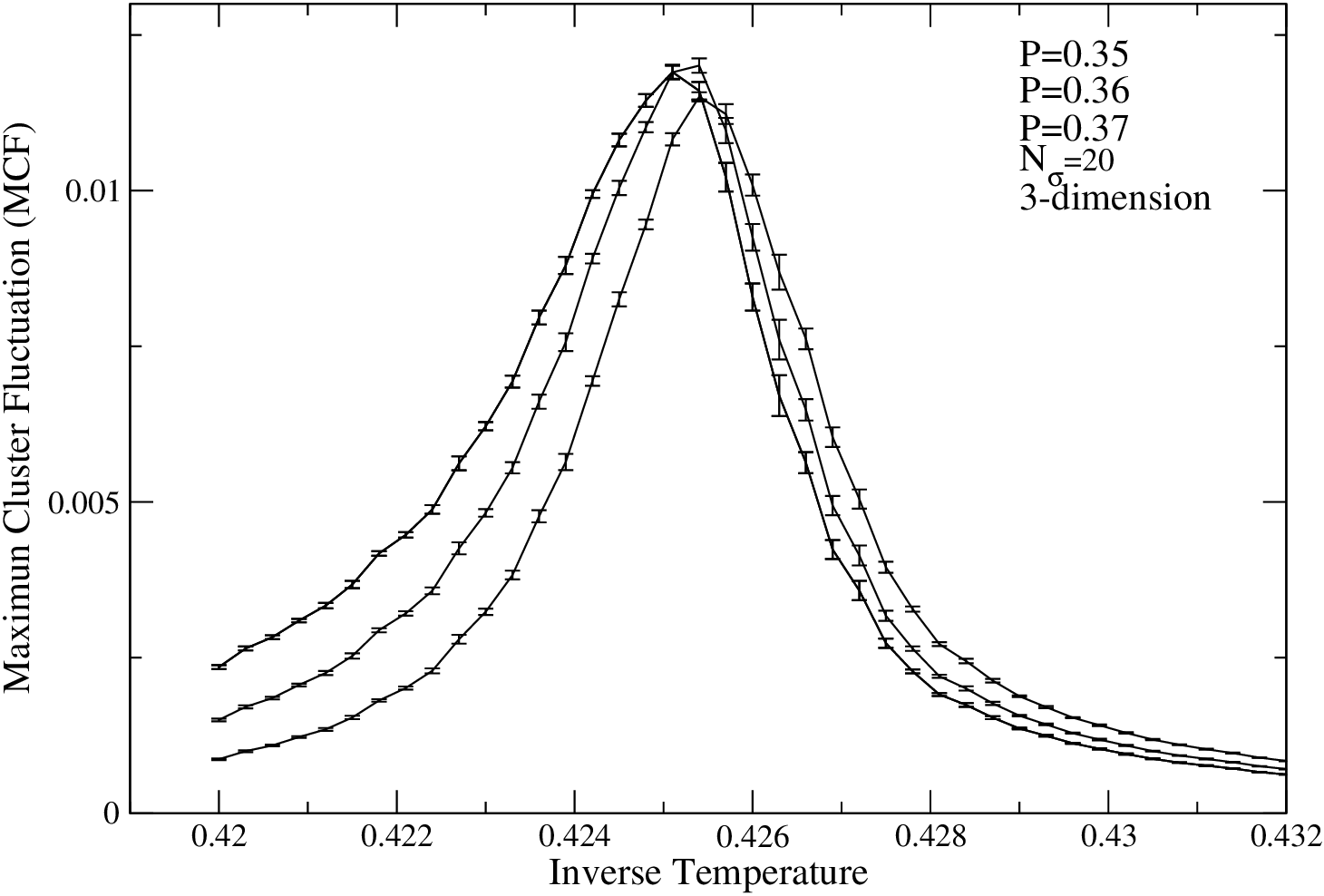}}
\caption{}
\label{fig1}
\end{figure}

\begin{figure}
\centering
\subfigure[]{\includegraphics[angle=0,width=10truecm]{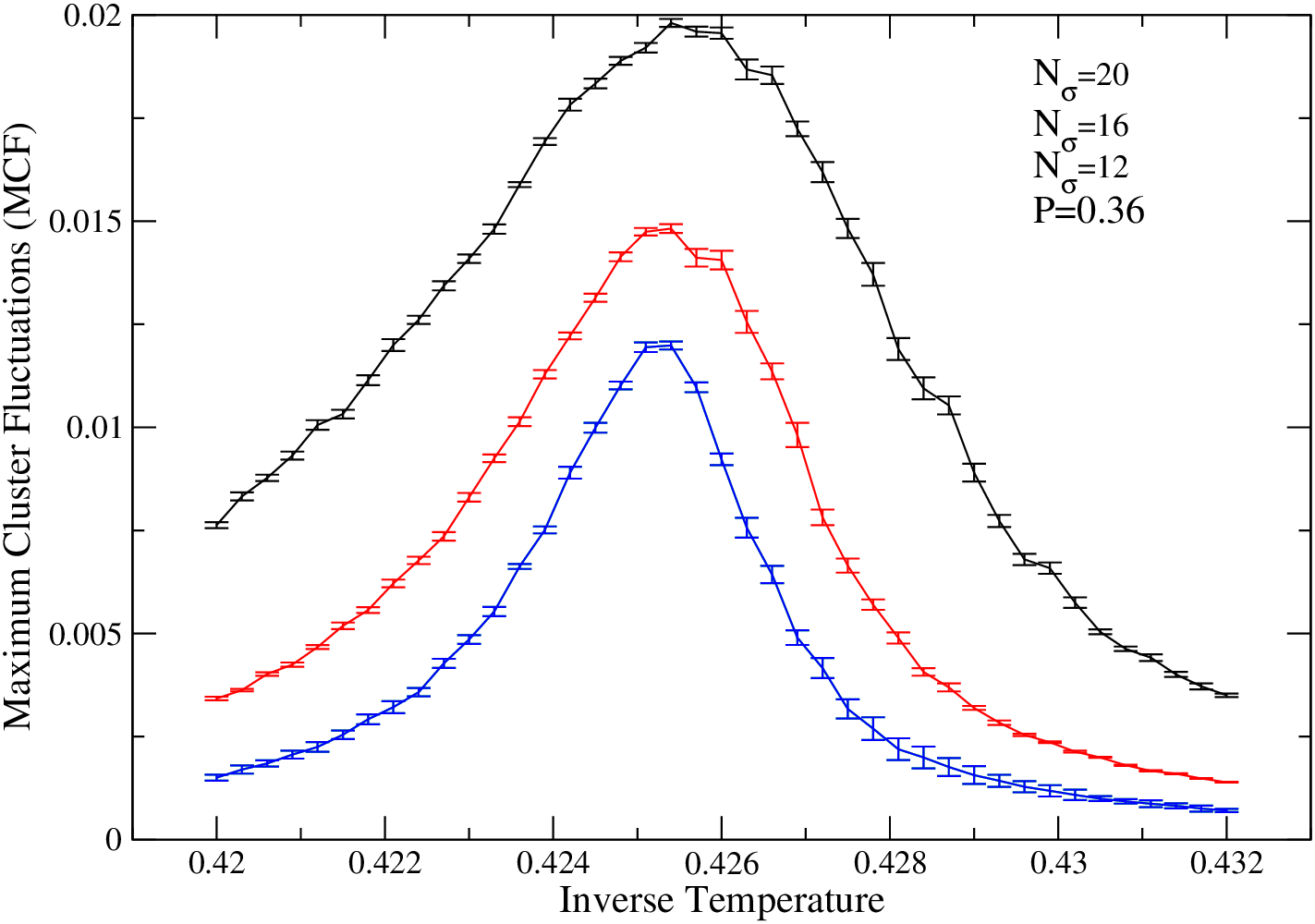}}\\
\subfigure[]{\includegraphics[angle=0,width=10truecm]{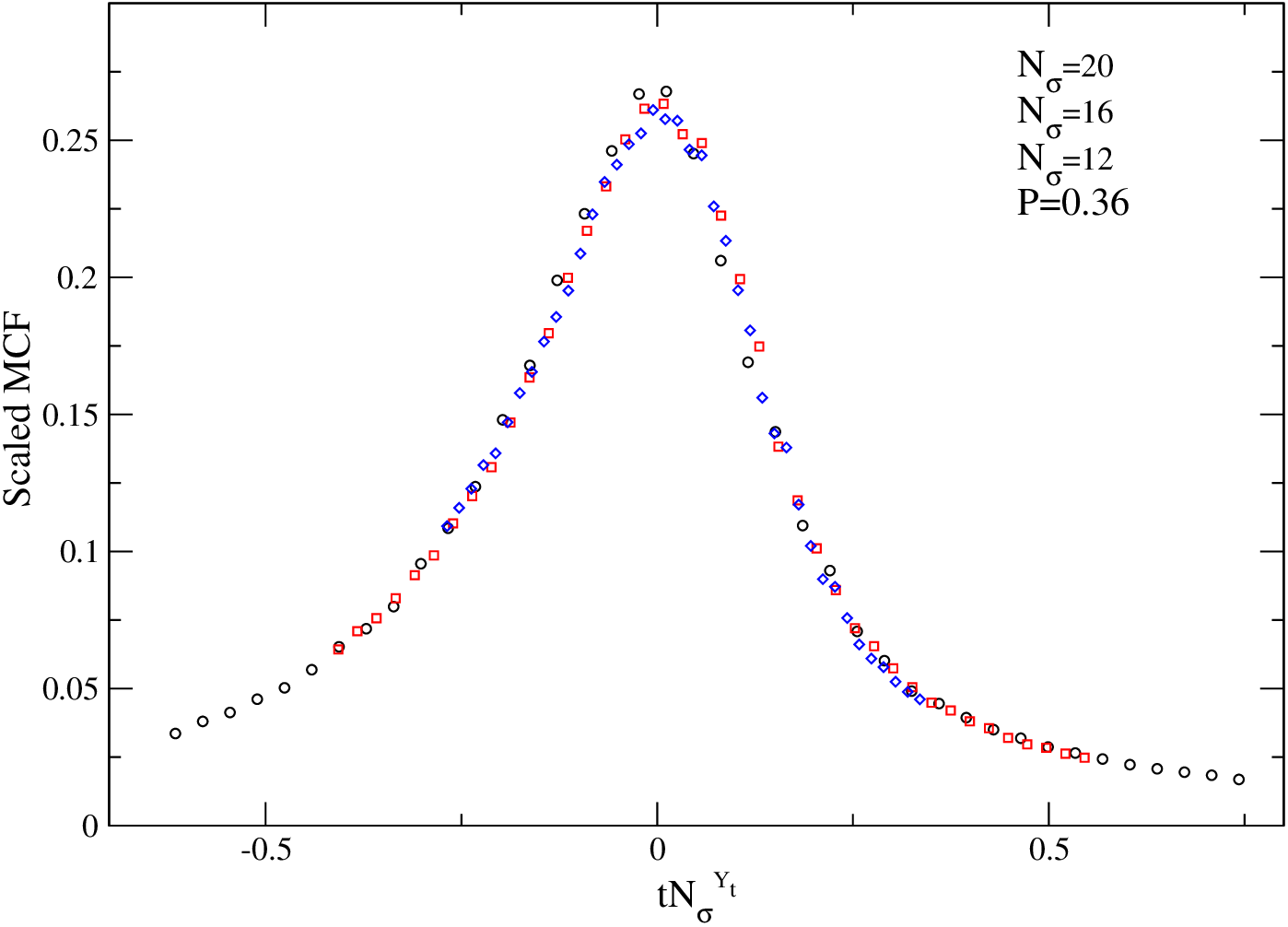}}
\caption{}
\label{fig2}
\end{figure}

\begin{figure}
\centering
\subfigure[]{\includegraphics[angle=0,width=10truecm]{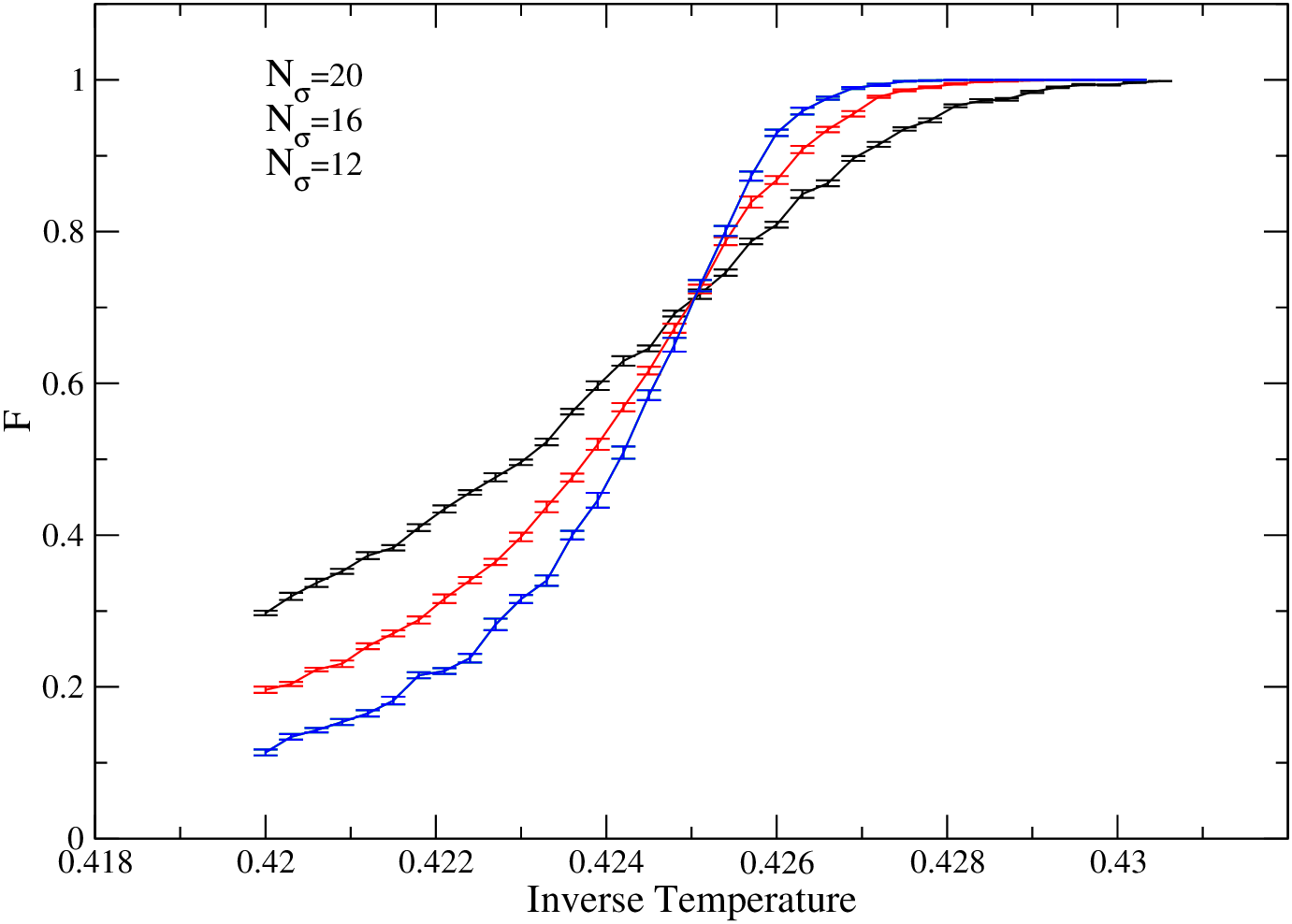}}\\
\subfigure[]{\includegraphics[angle=0,width=10truecm]{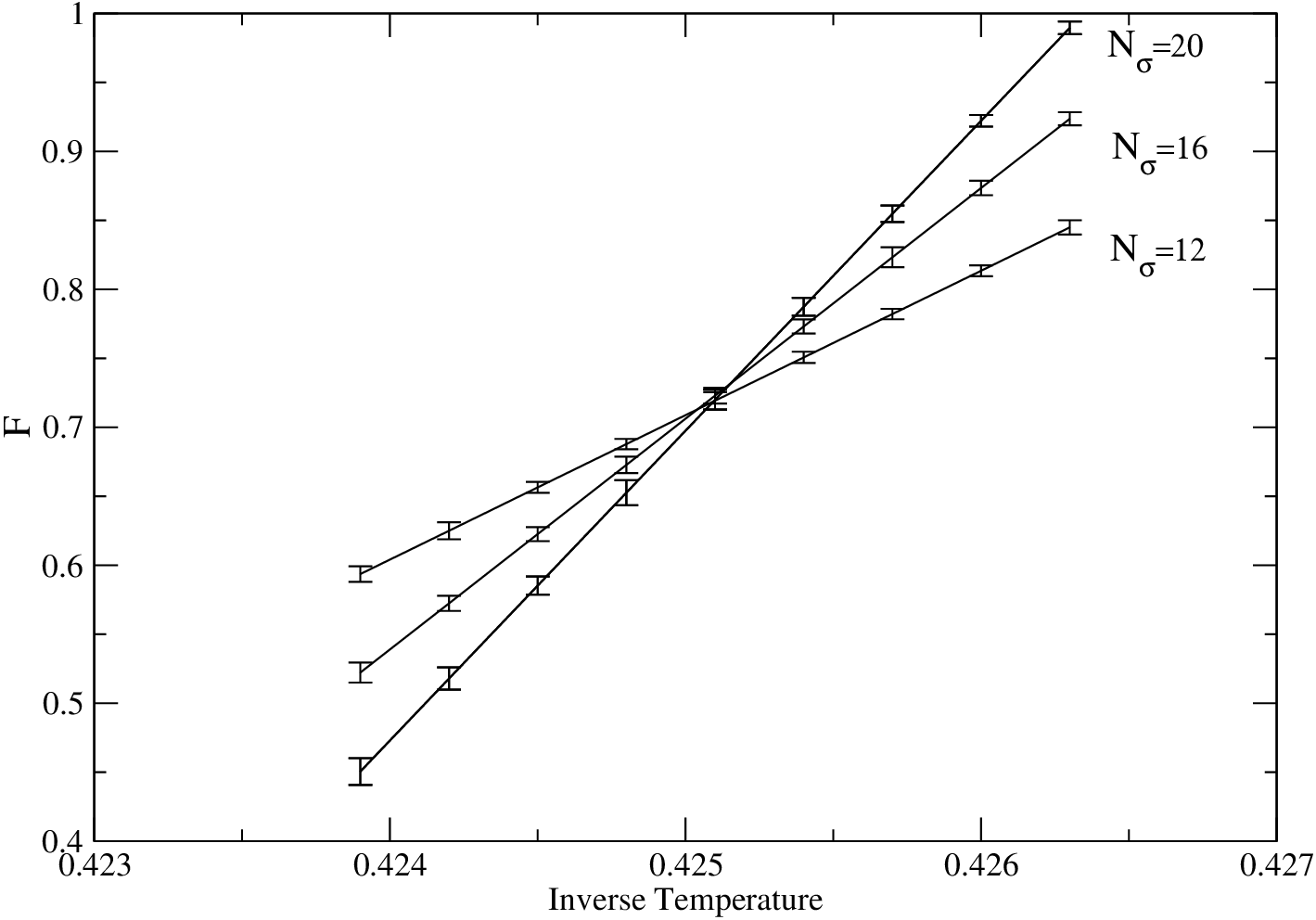}}
\caption{}
\label{fig3}
\end{figure}


\begin{thebibliography}{99}

\bibitem{Polyakov-Susskind} A. M. Polyakov, Physics Letters  {\bf B 72} (1978) 477 ;
L. Susskind, Physical Review  {\bf D  20} (1979) 2619 .


\bibitem{SvetitskyYaffe} B. Svetitsky, L. Yaffe,
Nuclear Physics  {\bf B 210} [FS6] (1982) 423 ; 
L. Yaffe, B. Svetitsky, Physical Review   {\bf D 26} (1982) 963; 
B. Svetitsky, Physics Reports {\bf 132} (1986) 1 .

\bibitem{Kertesz} J. Kert\'esz, Physica  {\bf A 161} (1989) 58 .



\bibitem{Satz1} H. Satz, Nucl. Phys. {\bf A 642} (1998) 130 .

\bibitem{Satz2} H. Satz, Nucl. Phys. {\bf A 681} (2001) 3.

\bibitem{KasteleynFortuin}P. W. Kasteleyn, C. M. Fortuin, Journal of the
 Physical Society of Japan {\bf 26} (Suppl.) (1969) 11 .


\bibitem{ConiglioKlein} A. Coniglio, W. Klein, Journal of Physics  {\bf A 13} (1980) 2775 .

\bibitem{Fort1} S. Fortunato, F. Karsch, P. Petreczky, H. Satz, Phys. Lett. {\bf B502} (2001) 321

\bibitem{Fort2} S. Fortunato, F. Karsch, P. Petreczky, H. Satz, 
Nucl. Phys. Proc. Suppl. {\bf 94} (2001) 398 .

\bibitem{Fort3}S. Fortunato and H. Satz, Nucl. Phys. {\bf A 681} (2001) 466 .

\bibitem{Fort4} S. Fortunato, H. Satz, Nucl. Phys. Proc. Suppl. {\bf 83} (2000) 452 .

\bibitem{Blanchard}P. Blanchard , S. Digal , S. Fortunato, D. Gandolfo, T. Mendes, H. Satz
J.Phys. {\bf A 33} (2000) 8603 .

\bibitem{Fort5}S. Fortunato, H. Satz, Nucl. Phys. {\bf B 598} (2001) 601 . 

\bibitem{Bialas}P. Bialas, P. Blanchard, S. Fortunato, D. Gandolfo, 
H. Satz, Nucl. Phys. {\bf B 583} (2000) 368 .

\bibitem{GreenKarsch} F. Green, F. Karsch, Nucl. Phys.  {\bf B 238} (1984) 297 .

\bibitem{StaufferAharony}D. Stauffer, A. Aharony, Taylor {\&} Francis, London 1994.


\bibitem{semraprob} S. Demirt\"urk, Y. G\"und\"u\c{c}, International Journal of Modern Physics {\bf C 12} (2001) 1361.

\bibitem{oliveira1} P.M.C. de Oliveira, Europhys. Lett. {\bf 20}, (1992) 621.

\bibitem{oliveira2} P.M.C. de Oliveira, Physica {\bf A 205}, (1994) 101.

\bibitem{oliveira3} J.M. de F. Neto, S.M. de Oliveira, and P.M.C. de Oliveira, Physica {\bf A 206}, (1994) 463.

\bibitem{oliveira4} P.M.C. de Oliveira, S.M. de Oliveira, C.E. Cordeiro, and D. Stauffer, J. Stat. Phys. {\bf 80}, (1995) 1433.


\bibitem{semrascaling} S. Demirt\"urk, N. Seferoglu, M. Ayd{\i}n and Y. G\"und\"u\c{c},  
International Journal of Modern Physics {\bf C 12},  (2001) 403.

\end{thebibliography}
\end{document}